\def\e{{\rm e}}

\def\half{{1\over2}}

\def\abs#1{{\left|{#1}\right|}}

\def\st{{\tilde t}}
\def\gtsim{\mathrel{\hbox{\raise0.2ex
\hbox{$>$}\kern-0.75em\raise-0.9ex\hbox{$\sim$}}}}
\def\ltsim{\mathrel{\hbox{\raise0.2ex
\hbox{$<$}\kern-0.75em\raise-0.9ex\hbox{$\sim$}}}}
\newcommand{\PRD}[3]{Phys. Rev. {\bf D{#1}} (19{#2}) {#3}}

\newcommand{\NPB}[3]{Nucl. Phys. {\bf B{#1}} (19{#2}) {#3}}
\newcommand{\PLB}[3]{Phys. Lett. {\bf B{#1}} (19{#2}) {#3}}
\newcommand{\PTP}[3]{Prog. Theor. Phys. {\bf {#1}} (19{#2}) {#3}}

\documentclass[12pt]{article}
\usepackage{epsf}
\usepackage{latexsym}
\usepackage{bm}
\makeatletter

\@addtoreset{equation}{section}
\makeatother
\begin{document}
\begin{titlepage}
\begin{flushright}
SAGA--HE--146\\
\date={March 6, 1999}
\end{flushright}
\vspace{24pt}
\centerline{\Large {\bf Transitional $CP$ Violation in MSSM and}}
\vspace{8pt}
\centerline{\Large {\bf Electroweak Baryogenesis}}
\vspace{8pt}
\vspace{24pt}
\begin{center}
{\bf Koichi~Funakubo$^{a,}$\footnote{e-mail: funakubo@cc.saga-u.ac.jp},
Shoichiro~Otsuki$^{b,}$\footnote{e-mail: otks1scp@mbox.nc.kyushu-u.ac.jp}
and Fumihiko~Toyoda$^{b,}$\footnote{e-mail: ftoyoda@fuk.kindai.ac.jp}}
\end{center}
\vskip 0.8 cm
\begin{center}
{\it $^{a)}$Department of Physics, Saga University,
Saga 840-8502 Japan}
\vskip 0.2 cm
{\it $^{b)}$Kyushu School of Engineering, Kinki University, 
Iizuka 820-8255 Japan}
\end{center}
\baselineskip=20pt
\vskip 1.0 cm
\bigskip\bigskip
\abstract{
Electroweak baryogenesis depends on the profile of the  
bubble wall created in the first-order phase transition.
It is pointed out that $CP$ violation in the Higgs sector
of the MSSM could  become large enough to explain the baryon asymmetry.
We confirm this by solving the equations of motion for the Higgs 
fields with the effective potential at the transition temperature.
That is, we present an example such that the transitional $CP$ violation is
realized and show the possibility that the baryon asymmetry of the universe
may be produced, if marginally, by the $\tau$ 
lepton interacting with the wall, when an explicit $CP$ breaking 
in the Higgs sector, which is consistent with experimental bounds,
is induced  at the phase transition.         
}
\vfill\eject
\setcounter{footnote}{0}
\end{titlepage}
%
%
%%%%%%%%%%%%%%%%%%%%%%%%%%%%%%%%%%%%%%
\section{Introduction}
Baryogenesis\cite{reviewEB} at the electroweak phase transition (EWPT) is 
an attractive mechanism to explain the baryon asymmetry of the universe (BAU) 
because the mechanism depends on  parameters that will be tested sooner 
or later on the earth.
For the mechanism to be viable, however, some extension of the standard model 
is required to guarantee first-order EWPT and to incorporate other sources 
of $CP$ breaking than the CKM phase.\par
Among possible extensions of the standard model,
the minimal supersymmetric standard model (MSSM) may be a well-motivated one.
The MSSM could cause a strongly first-order phase transition when one of the 
stops is light\cite{CQW,DGG} and has some sources of $CP$ breaking,
though they are constrained by measurements of the 
neutron electric dipole moment (EDM)\cite{Dugan,KO}.  
On the other hand, another source of $CP$ violation  
is relative phase $\theta$ of the vacuum expectation values (VEVs) of 
the two Higgs doublets\cite{NKC}. 
The Higgs VEVs including the phase, which characterize an expanding
bubble wall created at the first-order EWPT, vary spatially. 
Such $CP$ violation affects propagation of quarks and leptons through
their Yukawa couplings and that of charginos, neutralinos and 
sfermions through their mass matrices, so that weak hypercharge is 
carried by these particles into the symmetric phase region, where
the hypercharge will be turned into baryon number by sphaleron
processes.\par
In previous papers\cite{FKOTTb,FKOTe}, we attempted to determine the 
profile of the bubble wall by solving equations of motion for an effective 
potential $V_{\rm eff}$ at the transition temperature $T_{C}$ in the 
two-Higgs-doublet model under some discrete symmetry.
For a phenomenological set of the effective parameters, we presented a 
solution such that the $CP$-violating phase $\theta$
spontaneously generated becomes as large as $O(1)$ around the wall,
while it completely vanishes in the broken and symmetric vacua.
We shall refer to this mechanism as {\it{transitional $CP$ violation}}.
This solution yields the hypercharge flux by 
the quark or lepton transport to generate the BAU.
We also showed \cite{FKOTb} that a possible explicit $CP$ breaking
in the Higgs sector at $T_{C}$ does {\it nonperturbatively} resolve
the degeneracy between the $CP$-conjugate pair 
of the bubbles and leave a certain amount of BAU after the EWPT.
Although several necessary conditions for the transitional $CP$
violation were found in the MSSM\cite{FKOTd}\footnote{As for studies
including nonperturbative effects by use of 3d effective theory,
see \cite{Laine}}, we must solve the 
equations for the wall profile in order to confirm that such a 
mechanism works in practice.
% \footnote{.
% By a calculation in the MSSM\cite{HN},  the $CP$  
% breaking responsible for the BAU may give the neutron EDM larger by
%  two orders of magnitude than the experimental bound.
%  }.
\par
In the present article, we investigate possibility of the transitional
$CP$ violation in the MSSM. First of all in \S 2, we study the effective
potential at $T_{C}$ following a method proposed by one of 
the authors(K.~F.)\cite{F}.
It is approximated by polynomials in the Higgs fields whose coefficients
are given by the effective parameters\cite{FKOTd,Comelli,EMQ}.
In particular, we also point out that an explicit $CP$ breaking in the
Higgs sector, which is induced through loop corrections from the
SUSY particles, could be enhanced at the EWPT.
Employing the approximated effective potential, the equations of
motion for classical field configurations are derived to find a 
$CP$-violating profile of a bubble wall in \S 3.
In \S 4, we give one of numerical examples of the effective parameters 
which, at zero temperature, produce mass spectrum consistent with
present observations.
The global structure of the effective potential suggests that two 
degenerate minima of it are connected by a path with almost constant 
$\tan\beta$, where $\tan\beta$ is the ratio of the expectation values
of the two Higgs doublets\footnote{It is also reported that the spatial
dependence of $\beta$ is very small: $\Delta\beta \simeq O(10^{-3})$\cite{MQS}.}.
This justifies to neglect spatial dependence of $\tan\beta$ so that
the number of unknown functions is reduced.
Though the bubble wall is rather broad,  transitionally 
$CP$-violating solutions by the effective parameters may 
produce the BAU by the $\tau$ lepton transport, if marginally, 
when the explicit $CP$ breaking in the Higgs sector is induced from
complex parameters consistent with the neutron EDM measurement.
A summary and discussions are given in \S 5. 
Formulae for the effective parameters are given in  Appendix A. 
%%%%%%%%%%%%%%%%%%%%%%%%%%%%%%%%
%%%%%%%%%%%%%%%%%%%%%%%%%%%%%%%%%%
\section{Effective Potential and Transitional $CP$ Violation}
Let us parameterize the Higgs doublets of the MSSM as  
\begin{equation}
  \varphi_d
 ={1\over{\sqrt2}}\left(\begin{array}{c} \rho_1 \\
                        0
        \end{array}\right)
 ={1\over{\sqrt2}}\left(\begin{array}{c} v_1 \\
                        0
        \end{array}\right),         \nonumber \\
  \varphi_u
 ={1\over{\sqrt2}}\left(\begin{array}{c} 0 \\
                        \rho_2\e^{i\theta}
        \end{array}\right)
 ={1\over{\sqrt2}}\left(\begin{array}{c} 0 \\
                        v_2+iv_3
        \end{array}\right),
	\label{eq:VEV2}
\end{equation}
\par\noindent
where $\theta\equiv\theta_1-\theta_2$. The most general potential for them 
at the tree level, which is gauge invariant and renormalizable, is given as
\begin{eqnarray}
 V_0 &=&
 m_1^2\varphi_d^\dagger\varphi_d + m_2^2\varphi_u^\dagger\varphi_u
 +(m_3^2\varphi_u\varphi_d + \mbox{h.c})     \nonumber\\
 & &
 +{{\lambda_1}\over2}(\varphi_d^\dagger\varphi_d)^2
 +{{\lambda_2}\over2}(\varphi_u^\dagger\varphi_u)^2
 +\lambda_3(\varphi_u^\dagger\varphi_u)(\varphi_d^\dagger\varphi_d)
 +\lambda_4(\varphi_u\varphi_d)(\varphi_u\varphi_d)^* \nonumber\\
 & &
 +\left[
  {{\lambda_5}\over2}(\varphi_u\varphi_d)^2 +
  (\lambda_6\varphi_d^\dagger\varphi_d+\lambda_7\varphi_u^\dagger\varphi_u)
  \varphi_u\varphi_d + \mbox{h.c}
  \right].
\end{eqnarray}
\par\noindent
Here the coefficients are as follows:
\begin{eqnarray}
 m_1^2 &=& \tilde{m}_d^2+|\mu|^2,\qquad
 m_2^2  =  \tilde{m}_u^2+|\mu|^2,\qquad
 m_3^2  =  \mu B,   \nonumber\\
 \lambda_1&=&\lambda_2={1\over4}(g_2^2+g_1^2),\qquad
 \lambda_3 = {1\over4}(g_2^2-g_1^2),\qquad
 \lambda_4 = -\half g_2^2,    \\
 \lambda_5 &=& \lambda_6 = \lambda_7 = 0,
\end{eqnarray}
where $g_{2(1)}$ is the $SU(2)(U(1))$ gauge coupling and $\mu$ is the 
coefficient of the Higgs quadratic interaction in the superpotential.
The mass squared parameters $\tilde{m}_u^2$, $\tilde{m}_d^2$ and $\mu B$
come from the soft SUSY-breaking terms so that they are arbitrary at this
level. $m_3^2$ could be complex but its phase can be eliminated by rephasing 
the fields when $\lambda_5=\lambda_6=\lambda_7=0$.
We adopt the convention in which this $m_3^2$ is real and positive.
\par
%%%%%%%%%%%%%%%%%%%%%%%%%%%%%%%%%%%
\subsection{Effective potential}
The effective potential up to the one-loop terms of radiative and 
finite-temperature corrections is
\begin{equation}
 V_{\rm eff}=V_0+V_1(\rho_i,0)+V_1(\rho_i,T),
\end{equation}
where
\begin{equation}
 V_1(\rho_i,0)=\sum_{j} {n_j\over64\pi^2}m_j^4
 \left[\log\left({m_j^2 \over M_{\rm ren}^2}\right)-{3\over2}\right],
\end{equation}
\begin{equation}
 V_1(\rho_i,T) =
 {T^4\over2\pi^2}\sum_j n_j J_{\pm}\left(a_j^2={m_j^2\over T^2}\right),
\end{equation}
with
\begin{equation}
 J_{\pm}(a_j^2) =
 \int_0^\infty dx\,x^2\log\left(1\pm \exp[-\sqrt{x^2+a_j^2}]\right).
                \label{eq:def-J}
\end{equation}
Here we  consider the contributions of gauge bosons, top quark, 
top squarks ($\tilde t$), charginos($\chi^{\pm}$) and 
neutralinos ($\chi^0$).
The $n_i$ counts the degrees of freedom of each species including its
statistics, that is, $n_j>0 (n_j<0)$ for bosons (fermions).
The $m_j$, which is a function of the Higgs background ($\rho_i,\theta$),
is the mass of the particle.
The mass matrices of the charginos, neutralinos and stops are given 
in  Appendix\footnote{We have  used the $\overline{\rm DR}$-scheme to 
renormalize $V_{\rm eff}$.
}.\par
We assume  $M_1=M_2$ for the gaugino mass parameters in the mass 
matrices, which simplifies the neutralino contributions so as to be
proportional to the chargino contributions.
The coefficients in $V_{0}$  are replaced by 
the corresponding effective parameters in $V_{\rm eff}$. 
Such effective parameters that are relevant to $CP$ violation in the 
Higgs sector are defined by
\begin{eqnarray}
 {\bar m_3^2} &=&
-\left.{{\partial^2V_{\rm eff}}\over{\partial v_1\partial v_2}}\right|_0=
 m_3^2 + \Delta_{\chi}m_3^2+ \Delta_{\tilde t}m_3^2 ,    \\
 {\bar\lambda}_5 &=&{1\over2}\left(\left.
{{\partial^4V_{\rm eff}}\over{\partial v_1^2\partial v_2^2}}\right|_0
-\left.{{\partial^4V_{\rm eff}}\over{\partial v_1^2\partial v_3^2}}\right|_0
 \right)=
 \Delta_{\chi}\lambda_5 + \Delta_{\tilde t}\lambda_5 ,   \\ 
 {\bar\lambda}_6 &=& 
-{1\over3}\left.{{\partial^4V_{\rm eff}}\over{\partial v_1^3\partial v_2}}
\right|_0
=\Delta_{\chi}\lambda_6 + \Delta_{\tilde t}\lambda_6 ,   \\ 
 {\bar\lambda}_7 &=& 
-{1\over3}\left.{{\partial^4V_{\rm eff}}\over{\partial v_1\partial v_2^3}}
\right|_0
=\Delta_{\chi}\lambda_7 + \Delta_{\tilde t}\lambda_7 , 
\end{eqnarray}
\par\noindent
where $\Delta_{\chi}$ ($\Delta_{\tilde t}$) implies the one-loop corrections 
from charginos and neutralinos (stops) given in  Appendix.
Except for the light stop ($\st_1$) contributions, Taylor expansion of the 
finite-temperature integrals around $\rho_i=0$ is valid.
As for the light stop, we employ the high-temperature expansion\cite{DJ}
to obtain
\begin{eqnarray}
 J_-(a_{\st_1}^2)&=& -{\pi^4\over45}+{\pi^2\over12}a_{\st_1}^2(\rho)
 -{\pi\over6}a_{\st_1}^3(\rho)+\lambda_- a_{\st_1}^4(\rho)+\ldots.
               \label{eq:high-T-exp-J}
\end{eqnarray}
where
  $ a_{\st_1}^2(\rho)={m_{\st_1}^2(\rho)/T^2}$ and 
$\lambda_-=0.1764974$.
\par
Finally, the effective potential up to the $\rho^{4}$-terms is 
expressed as
\begin{eqnarray}
 V_{\rm eff}(\rho_i,\theta_i)&=&
 \half {\bar m}_1^2\rho_1^2+\half {\bar m}_2^2\rho_2^2 - {\bar m}_3^2
\rho_1\rho_2\cos\theta
 +{{{\bar\lambda}_1}\over8}\rho_1^4+{{{\bar\lambda}_2}\over8}\rho_2^4   
\nonumber  \\
&+&{{{\bar\lambda}_3+{\bar\lambda}_4}\over4}\rho_1^2\rho_2^2
 + {{{\bar\lambda}_5}\over4}\rho_1^2\rho_2^2\cos2\theta
 - \half({\bar\lambda}_6\rho_1^2+{\bar\lambda}_7\rho_2^2)\rho_1\rho_2\cos\theta
    \nonumber  \\
&-& [A\rho_1^3+\rho_1^2\rho_2(B_0+B_1\cos\theta+B_2\cos2\theta)  
    \nonumber  \\
& & + \rho_1\rho_2^2(C_0+C_1\cos\theta+C_2\cos2\theta) + D\rho_2^3 ],
\label{eqn:effv}
\end{eqnarray}
where ${\bar m}_i$ and ${\bar\lambda}_i$ are the effective parameters, 
some of which are given above.  
The $\theta$-dependent $\rho^{3}$-terms come from the light stop 
contributions by the high-temperature expansion\cite{FKOTd}, while
the weak gauge bosons contribute to $\theta$-independent $\rho^3$-terms.
Note also that non-zero values of ${\bar\lambda}_5$, ${\bar\lambda}_6$ and 
${\bar\lambda}_7$ are induced.\par
To calculate the effective potential, we give 
$v_0=246\mbox{GeV}$ and $\tan\beta_0$ at zero temperature as an input.
Once $m_3^2$ at the tree level is given, the other mass parameters
in the tree-level potential is fixed by minimizing the effective 
potential at $T=0$\cite{F}. At the same time, the mass of the
lighter (heavier) $CP$-even Higgs scalar $m_h$ ($m_H$) and that of
the pseudoscalar $m_A$ are evaluated from the second derivatives of
the effective potential at the minimum.
When the EWPT is of first order, the transition temperature $T_C$
is determined as temperature at which $(\rho_1,\rho_2)=(0,0)$ and 
$(\rho_1,\rho_2)=(v\cos\beta,v\sin\beta)$ become degenerate minima
of the effective potential, where the symmetry-breaking minimum
is searched for by a numerical method explained in \cite{F}.
This can easily be extended to the case with $CP$ violation,
in which $\theta$ acquires nonzero value and the Higgs scalars and
pseudoscalar mix to produce new mass eigenstates\cite{F,Pilaftsis}.\par
%%%%%%%%%%%%%%%%%%%%%%%%%%%%%%%%%%%%%%
\subsection{Spontaneous $CP$ violation}
For simplicity, we assume that all the parameters of $V_{\rm eff}$
are real, that is, no explicit $CP$ breaking. 
Then at zero temperature $CP$ symmetry can be spontaneously violated 
but the lightest scalar becomes too light to satisfy
experimental lower bound\cite{M}.\par
Now we argue that if the EWPT is of first order, $CP$ can be violated
for broader range of acceptable parameters than at $T=0$.
Note that
\begin{equation}
 V_{\rm eff}(\rho_i,\theta_i) =
 F(\rho_1,\rho_2)\left[ \cos\theta-G(\rho_1,\rho_2)\right]^2 +
 \mbox{$\theta$-independent terms},
\end{equation}
where
\begin{eqnarray}
 F(\rho_1,\rho_2) &\equiv&
 {{\bar\lambda_5}\over2}\rho_1^2\rho_2^2 -
 2(B_2\rho_1^2\rho_2+C_2\rho_1\rho_2^2),     \label{eq:def-F}\\
 G(\rho_1,\rho_2) &\equiv&
 {{2{\bar m}_3^2+\bar\lambda_6\rho_1^2+\bar\lambda_7\rho_2^2+
   2(B_1\rho_1+C_1\rho_2)}\over
  {2\bar\lambda_5\rho_1\rho_2-8(B_2\rho_1+C_2\rho_2)}}.
                                             \label{eq:def-G}
\end{eqnarray}
Recall that spontaneous $CP$ violation occurs at $T=0$ ($B_i=C_i=0$), 
only if 
\begin{equation}
 F(\rho_1,\rho_2)>0 \qquad\mbox{and}\qquad -1<G(\rho_1,\rho_2)<1,
                           \label{eq:cond-sCPV}
\end{equation}
for $(\rho_1,\rho_2)=(v_0\cos\beta_0,v_0\sin\beta_0)$. These conditions
strictly restrict the range of $m_3^2$ so that a very light scalar
inevitably results, since ${\bar m}_3^2$ must be the same order as
$\bar\lambda_6\rho_1^2+\bar\lambda_7\rho_2^2$.
At the first-order EWPT, $(\rho_1,\rho_2)$ varies from
$(v\cos\beta,v\sin\beta)$ to $(0,0)$ between the broken and symmetric
phase regions. 
Then the conditions (\ref{eq:cond-sCPV}) could be satisfied
for broader range of parameters, which are acceptable in contrast to
the zero-temperature case, since the effective parameters receive
finite-temperature corrections and $(\rho_1,\rho_2)$ is not specified
to a single point. In this case, there exists a 
$CP$-violating local minimum in the transient region and the bubble
wall profile, which is represented by the classical Higgs 
configuration, could have nontrivial $\theta$ near such a minimum.\par
Even when $F(\rho_1,\rho_2)<0$ for some $(\rho_1,\rho_2)$, 
the bubble wall profile could acquire nontrivial $CP$ phase as long as
$-1<G(\rho_1,\rho_2)<1$ is fulfilled at that point. 
In this case, $V_{\rm eff}$ for the fixed $(\rho_1,\rho_2)$
is convex upwards and its peak is located at $\theta\in(0,\pi)$.
If $G(\rho_1,\rho_2)$ varies between positive and negative values
along the wall profile, the minimum of $V_{\rm eff}$ for a fixed 
$(\rho_1,\rho_2)$ travels between $\theta=\pi$ and $\theta=0$ as
$\rho_i$ varies from the symmetric phase to the broken phase.
Then there exists a $CP$-violating saddle point of $V_{\rm eff}$
in the transient region.\par
In any case, to realize $-1<G(\rho_1,\rho_2)<1$ at some $(\rho_1,\rho_2)$,
$\abs{{\bar m}_3^2}$ at $T_C$ must become much smaller than tree-level 
$m_3^2$ by large negative corrections. 
Since the correction from the charginos (stops) to $m_3^2$ is proportional
to $\mu M_2$ ($\mu A_t$) as shown in Appendix, $\mu M_2<0$ and/or
$\mu A_t<0$ are required for transitional $CP$ violation to occur.
Once these parameters are provided, $m_3^2$ which leads to
$-1<G(\rho_1,\rho_2)<1$ is restricted in a rather narrow range.
Then the mass of the Higgs scalars and pseudoscalar are almost 
uniquely determined. 
Whether a transitionally $CP$-violating bubble wall is realized or not
is determined finally by solving the equations of motion with the effective 
potential at $T_C$.
%
%%%%%%%%%%%%%%%%%%%%%%%%%%%%%
\subsection{Explicit $CP$ breaking}
In the discussions before, all the parameters of $V_{\rm eff}$
are assumed to be real. However, 
an explicit $CP$ breaking is 
necessary to avoid the complete cancellation in the net baryon number 
due to the symmetry $\theta \leftrightarrow -\theta$ between the wall
profiles.\par
The origins of the explicit $CP$ breaking in $V_{\rm eff}$ are those in 
the complex $\mu$-parameter and SUSY-breaking parameters which 
depend on the phases,  $\alpha_1=\arg(\mu M_1)=\arg(\mu M_2)$  
contributed  from  charginos and neutralinos and $\alpha_2=\arg(\mu 
A_t^*)$ from   
 stops, as given in 
 Appendix. From these we have 
\begin{eqnarray}
 {\bar m}_3^2 &=& m_3^2+e^{i\alpha_1}\Delta_{\chi}^{(0)}m_3^2
                  +e^{i\alpha_2}\Delta_{\tilde t}^{(0)}m_3^2 , \\
 {\bar\lambda}_5 &=& e^{2i\alpha_1}\Delta_{\chi}^{(0)}\lambda_5
                  +e^{2i\alpha_2}\Delta_{\tilde t}^{(0)}\lambda_5 , \\
 {\bar\lambda}_6 &=& e^{i\alpha_1}\Delta_{\chi}^{(0)}\lambda_6
                  +e^{i\alpha_2}\Delta_{\tilde t}^{(0)}\lambda_6 , \\
 {\bar\lambda}_7 &=& e^{i\alpha_1}\Delta_{\chi}^{(0)}\lambda_7  
                  +e^{i\alpha_2}\Delta_{\tilde t}^{(0)}\lambda_7,
\end{eqnarray}
where $\Delta^{(0)}$ denotes the corrections in the case when all the 
tree-level parameters are real.
\par
The magnitudes of $\alpha_1$ and $\alpha_2$ will be bounded by  experiments 
to be $O(<10^{-3})$ or so\cite{Dugan}.
These $CP$-breaking phases can be gathered on ${\bar m}_3^2$ by rephasing 
the Higgs fields when the contributions of the stops are very small compared 
with those of charginos and neutralinos.
After rephasing, ${\bar m}_3^2$ is given in the form of
\begin{eqnarray}
  e^{-i\alpha_1}{\bar m}_3^2&=&e^{-i\alpha_1}m_3^2+\Delta_{\chi}^{(0)}m_3^2 
  \equiv e^{i\delta}|{\bar m}_3^2|,
\end{eqnarray}
so that the $CP$-breaking phase $\delta$ is given by 
\begin{eqnarray}
 \tan\delta=-{{m_3^2\sin\alpha_1}\over{m_3^2\cos\alpha_1+
 \Delta_{\chi}^{(0)}m_3^2}}.
\end{eqnarray}
This suggests a very interesting possibility that, 
if $\abs{m_3^2+\Delta_{\chi}^{(0)}m_3^2}\ll m_3^2$, which is often
the case for the transitional $CP$ violation, then a somewhat
large $|\delta|$ is induced even if $\sin\alpha_1\simeq 0$.
\par
Hereafter, we put the explicit $CP$ breaking at $T_C$ by replacing 
${\bar m_3^2}\cos\theta$ in $V_{\rm eff}$ (\ref{eqn:effv})
  by $(1/2)(e^{i(\delta+\theta)}
{\bar m_3^2}+{\rm h.c.})$, but consider the case of small $\delta\sim 
O(10^{-3})$. 
\par
%%%%%%%%%%%%%%%%%%%%%%%%%%%%%%%%%%
%%%%%%%%%%%%%%%%%%%%%%%%%%%%%%%%%%%
\section{$CP$-Violating Solutions}
%%%%%%%%%%%%%%%%%%%%%%%%%%%%%%
\subsection{Equations of motion}
We are now interested in  classical solutions of the bubble wall.
If the phase transition proceeds calmly, it will be valid to expect that 
the bubble wall grows keeping the profile of the critical bubble, which is 
determined by  static  equations of motion. Further, when the 
bubble is spherically symmetric and is sufficiently macroscopic so that it is 
regarded as a planar object, the system may be reduced to an effectively 
one-dimensional one.
Regarding the bubble wall as a static planar object, the equations of motion
 are given by\cite{FKOTTb}
\begin{eqnarray}
 {{d^2\rho_i(z)}\over{dz^2}}-\rho_i(z)\left({{d\theta_i(z)}\over{dz}}\right)^2
  -{{\partial V_{\rm eff}}\over{\partial\rho_i}} &=& 0,
	\label{eq:rho-z}  \nonumber\\
 {d\over{dz}}\left(\rho_i^2(z){{d\theta_i(z)}\over{dz}}\right)
  -{{\partial V_{\rm eff}}\over{\partial \theta_i}} &=& 0,
	\label{eq:theta-z}  
\end{eqnarray}
where $z$ is the coordinate perpendicular to the wall.
Furthermore, gauge configurations of the pure-gauge type with no dynamical 
freedom in 1+1-dimensions are expected to give the lowest energy of the system. 
Then, it is convenient to fix the gauge in such a way that the gauge
fields disappear by imposing the condition
\begin{equation} 
  \rho_1^2(z){{d\theta_1(z)}\over{dz}}+\rho_2^2(z){{d\theta_2(z)}\over{dz}}=0.
	\label{eq:sourceless-z}
\end{equation}
\par
The total energy density of the bubble wall per unit area is given by
\begin{equation}
  {\cal E}=\int_{-\infty}^\infty dz \left\{
  {1\over 2}\sum_{i=1,2}\left[\left({{d\rho_i}\over{dz}}\right)^2
    +\rho_i^2 \left({{d\theta_i}\over{dz}}\right)^2 \right]
   + V_{\rm eff}(\rho_1,\rho_2,\theta) \right\}.    
\label{eq:energy-y}
\end{equation}\par
%%%%%%%%%%%%%%%%%%%%%%%%%%%%%%%%%%
\subsection{Kink ansatz}
In accord with the first-order phase transition, we adopt an 
ansatz that the kink configuration connecting the vacua is a solution to the 
equations of motion. 
Using a dimensionless variable  $y=(1/2)(1-\tanh(az))$, 
we put $\rho(y)=(\rho_{1}/v)/\cos\beta=(\rho_{2}/v)/\sin\beta$ and
$\theta(y)=\theta_{1}/\sin^{2}\beta=-\theta_{2}/\cos^{2}\beta$ 
under the gauge-fixing condition (\ref{eq:sourceless-z}).
Here $a$ is taken to be the inverse wall thickness.  
The equations of motion to give  $\rho(y)$ and $\theta(y)$ are 
read as
\begin{eqnarray}
4y(1-y){d\over dy}\Bigl(y(1-y){d\rho\over dy}\Bigr)
-4\cos^{2}\beta\sin^{2}\beta y^{2}(1-y)^{2}\rho\Bigl({d\theta\over 
dy} \Bigr)^{2}
  &=&{\partial W(\rho,\theta)\over {\partial \rho} },  \label{eq:eom1} 
  \nonumber\\
  4\cos^{2}\beta\sin^{2}\beta y(1-y){d\over dy} 
  \Bigl(y(1-y)\rho^{2}{d\theta\over dy}\Bigr)
  &=&{\partial W(\rho,\theta)\over \partial\theta},  \label{eq:eom2}
\end{eqnarray}
where $W(\rho(y),\theta(y)) \equiv V_{\rm 
eff}(\rho_{i},\theta)/a^{2}v^{2}$ is dimensionless.\par
The parameterization of $W$ convenient to the kink ansatz in the 
presence of $\delta$ is as 
follows:
\begin{eqnarray}
W(\rho,\theta)& = &\rho^{2}\Bigl[ 
                 2+b\bigl(\cos(\theta_{b}+\delta)-\cos(\theta+\delta)\bigr)\Bigr]
                 \nonumber\\
              & + &\rho^{4}\Bigl[ 
                 2+c\bigl(\cos\theta_{b}-\cos\theta\bigr)+
                 {d\over4}\bigl(\cos2\theta_{b}-\cos2\theta\bigr)\Bigr]
                 \nonumber\\
              & - &\rho^{3}\Bigl[ 
                 4+e\bigl(\cos\theta_{b}-\cos\theta\bigr)+
                 f\bigl(\cos2\theta_{b}-\cos2\theta\bigr)\Bigr].
                 \label{eq:vdef2}
\end{eqnarray}
Here $\theta_{b}\sim O(\delta)$ is the boundary value of $\theta$ in 
the broken vacuum given later.
 The parameters in $W(\rho,\theta)$ and those in $V_{\rm eff}(\rho_{i},\theta)
$
are related as follows:
\begin{eqnarray}
 b &=& (1/a^{2}){\bar m}^{2}_{3} \cos\beta\sin\beta,\nonumber\\
 c &=& 
(v/a)^{2}({\bar \lambda}_{6}\cos^{2}\beta+{\bar \lambda}_{7}\sin^{2}\beta)\cos
\beta\sin\beta/2,\nonumber\\
 d &=& -(v/a)^{2}{\bar \lambda}_{5}\cos^{2}\beta\sin^{2}\beta,\nonumber\\
 e &=&  
 -(v/a^{2})(B_{1}\cos\beta+C_{1}\sin\beta)\cos\beta\sin\beta,\nonumber\\
 f &=&   
 -(v/a^{2})(B_{2}\cos\beta+C_{2}\sin\beta)\cos\beta\sin\beta,
   \label{eq:wparam1}
\end{eqnarray}  
together with relations from the kink ansatz: 
\begin{eqnarray}
 {\bar m}^{2}_{1}\cos^{2}\beta+{\bar m}^{2}_{2}\sin^{2}\beta  &=&
      2a^{2}(2+b\cos(\theta_{b}+\delta)),\nonumber\\
 {\bar \lambda}_{1}\cos^{4}\beta+{\bar \lambda}_{2}\sin^{4}\beta
   &+ & 2({\bar \lambda}_{3}+{\bar \lambda}_{4})\cos^{2}\beta \sin^{2}\beta\nonumber\\
    &=& (a/v)^{2}8(2+c\cos\theta_{b}+(d/4)\cos2\theta_{b}),\nonumber\\
    A\cos^{3}\beta &+&     
     (B_{0}\cos\beta+C_{0}\sin\beta)\cos\beta\sin\beta
    +D\sin^{3}\beta\nonumber\\
      &=&(a^{2}/v)(4+e\cos\theta_{b}+f\cos2\theta_{b}) \label{eq:wparam2}. 
\end{eqnarray}
Once the parameter set $(b,c,e,d,f)$ is given, the stability of the two vacua 
and
the condition $W(\rho,\theta) \geq 0$ in the 
region of $0\leq \rho<\infty$ and $0\leq\theta<2\pi$ 
have to be checked.
As it should be, in the case of $\delta=0$, the kink solution
 $(\rho=1-y\equiv\rho_{kink},\theta=0)$ satisfies the equations of 
 motion.\par
That ${\partial W}/{\partial \theta}|_{\rho=1}=0$ gives the boundary 
value $\theta_{b}$ determined from
\begin{equation}
 \tan\theta_{b} 
 =-\frac{b\sin\delta}{b\cos\delta+c-e+(d-4f)\cos\theta_{b}},         
\end{equation}
 and ${\partial W}/{\partial 
\theta}|_{\rho\simeq 0}=0$ does the boundary value of the symmetric 
vacuum as $\theta_{s}=-\delta,\pi-\delta$.
Needless to say, 
$\rho_b=1$ and $\rho_s=0$.
At first sight, any $\theta_{s}$ might be allowed because of 
$\rho_s=0$. But the energy density of the bubble wall diverges 
 for the other $\theta_s$.\par
  In the case of $\delta=0$,  suppose that we obtain a nontrivial solution 
$(\rho\ne\rho_{kink},\theta\ne 0)$ with the  maximum of $\theta>0$. Let us 
denote this solution as $(\rho^{0},\theta^{0})$. As explained before, we have 
the $CP$-conjugate partner of it,  
$(\rho^{0},-\theta^{0})$.
Without loss of generality we choose $\delta\geq 0$ because of the 
symmetry $(\delta,\theta)\longleftrightarrow(-\delta,-\theta)$.
For $\delta>0$ small enough, we would have in general three types of 
solutions, the positive-$\theta$ solution $(\rho^{+},\theta^{+})$, 
the negative-$\theta$ solution $(\rho^{-},\theta^{-})$  and the small-$\theta$ 
solution $(\rho^{s},\theta^{s})$ such that, as $\delta\rightarrow 0$, 
the positive-$\theta$ solution tends 
to $(\rho^{0},\theta^{0})$, the negative-$\theta$ solution to 
$(\rho^{0},-\theta^{0})$ and the small-$\theta$ solution to the kink solution 
$(\rho_{kink},\theta=0)$. The small-$\theta$ solution has no $CP$-conjugate
partner. For a larger $\delta$, either of the positive-$\theta$ 
solution or the negative-$\theta$ one,
which has the lower energy for a small $\delta$, survives\cite{FKOTb}.\par
The energy density of the bubble wall is now  expressed as follows, where
that of the kink solution, ${\cal E}_{kink}=av^{2}$/3, is chosen as 
the standard:
\begin{eqnarray}
  {\Delta{\cal E} \over av^{2}} & = & 
  \int_{0}^{1}dy\Biggl[y(1-y)\left\{\left({{d\rho}\over{dy}}\right)^2
              +\cos^2\beta\sin^2\beta\,\rho^2
               \left({{d\theta}\over{dy}}\right)^2\right\}\nonumber\\
    & \qquad +&{1 \over {2y(1-y)}}W(\rho(y),\theta(y))\Biggr]-{1\over3}.
    \label{eq:deleps}
\end{eqnarray}
Let us denote as $\Delta\cal E^{+}$, $\Delta\cal E^{-}$ and 
${\Delta\cal E}^s$ 
respectively for the positive-$\theta$, negative-$\theta$ and small-$\theta$
solutions. 
%%%%%%%%%%%%%%%%%%%%%%%%%%%%%%%%%%%%%%%%%%%%%%%%%%
%%%%%%%%%%%%%%%%%%%%%%%%%%%%%%%%%%%%%%%%%%%%%%%%%
\section{A Numerical Example of Transitional $CP$ Violation}
%%%%%%%%%%%%%%%%%%%%%%%%%%
\subsection{Numerical results of MSSM calculation}
We give one of numerical examples at the EWPT and the effective parameters
in the MSSM.
The parameters adopted are listed in Table~\ref{tab:1}.
As noted in \S 2.2, $\mu M_2<0$ and $\mu A_t<0$ are favored for
transitional $CP$ violation. Once these parameters are given,
the tree-level $m_3^2$ is tuned to satisfy the conditions for
transitional $CP$ violation. Since the magnitudes of the effective 
couplings at $T_C\sim100\mbox{ GeV}$ are always $10^{-3}-10^{-2}$
when the mass parameters are taken to be between weak scale and a few TeV, 
an appropriate value of $m_3^2$ is about $O(10^4)\mbox{ GeV}^2$.
Although it is reported that a small value of $\tan\beta_0$ is
favored for a strong EWPT satisfying $v/T_C>1$ with an acceptable 
Higgs mass\cite{E}, we take $\tan\beta_0=6$.
One of reasons why we take $\tan\beta_0=6$ is that for the other 
parameters listed in Table~\ref{tab:1}, a larger $\tan\beta_0$
yields a heavier Higgs scalar when one includes contributions
from charginos and neutralinos, which were often ignored in
literatures\cite{F}. 
Another reason is that for a large $\tan\beta_0$, it is rather
easy to find a stable nontrivial wall profile, since 
the second kinetic term in (\ref{eq:deleps}) $\sim \cos^{2}\beta$
while $W(\rho,\theta) \sim \cos\beta$ as in (\ref{eq:wparam1}),
so that ${\Delta \cal E}<0$ is realized.\par
\begin{figure}
 \epsfysize=6.0cm
 \centerline{\epsfbox{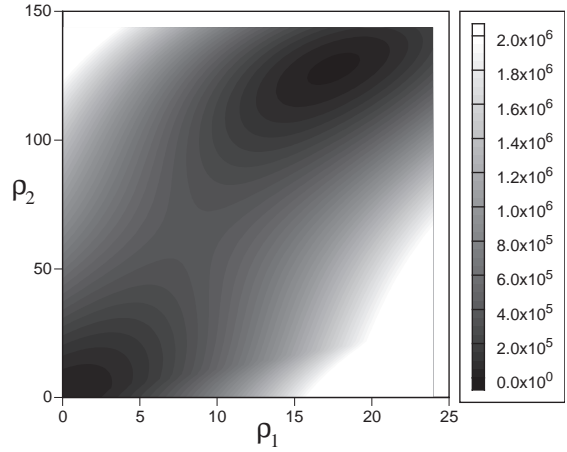}}
 \caption{Contour plot of $V_{\rm eff}(\rho_i;T_C)$ in the unit
 of $\mbox{GeV}^4$, where $\rho_1=v\cos\beta$ and $\rho_2=v\sin\beta$
 in the unit of $\mbox{GeV}$.}
 \label{fig:Veff-contour}
\end{figure}
\begin{figure}
 \epsfysize=5.0cm
 \centerline{\epsfbox{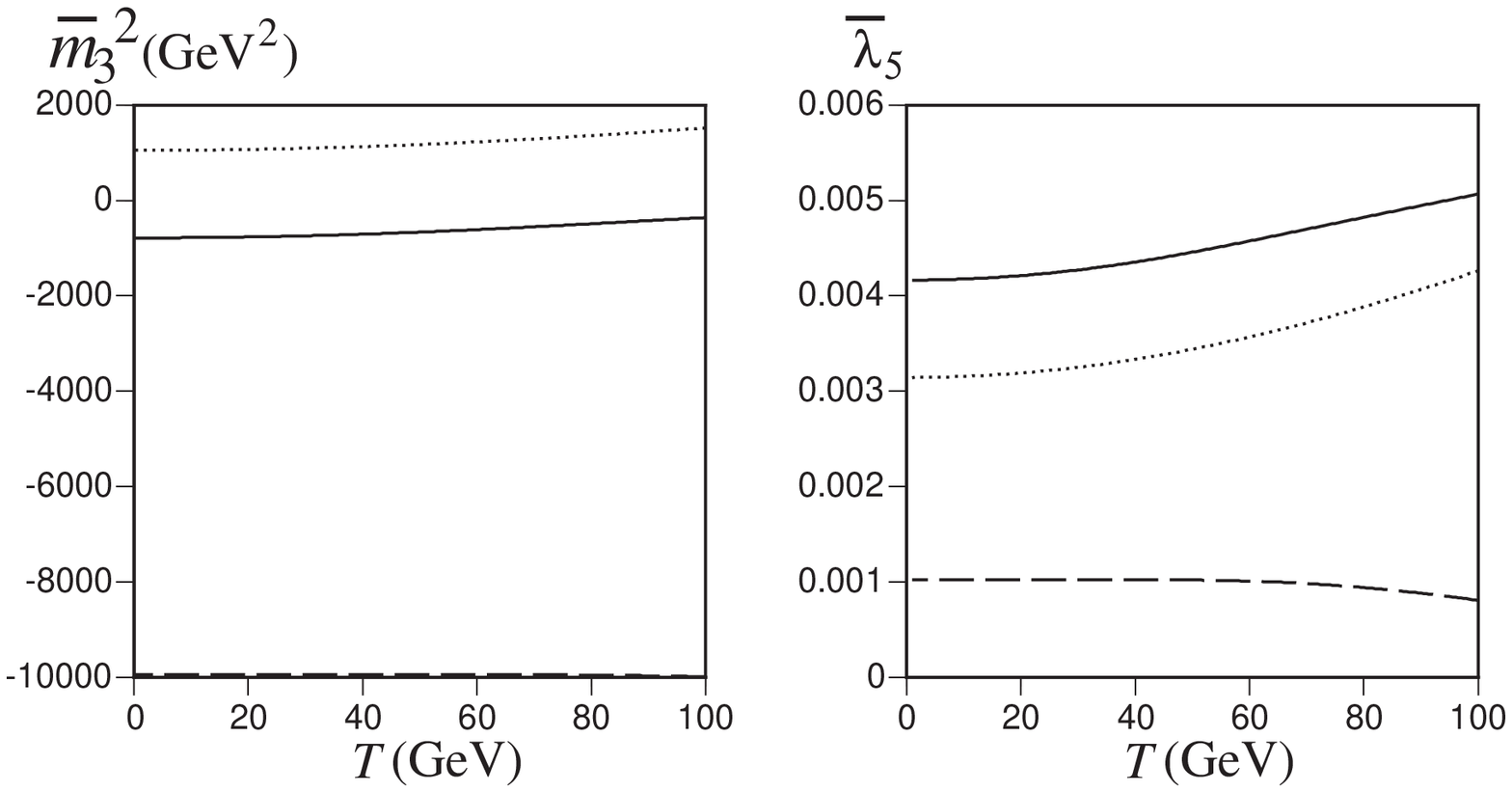}}
 \caption{Behaviors of the effective parameters 
 $\left({\bar m}_3^2\right)_{\rm eff}$
 and $\bar\lambda_5$. The dashed lines represent the contributions from 
 the charginos and neutralinos, the dotted lines represent those from 
 the stops and the solid lines are the sum of the tree-level and 
 one-loop contributions.}
 \label{fig:eff-para-1}
\end{figure}
\begin{figure}
 \epsfysize=5.0cm
 \centerline{\epsfbox{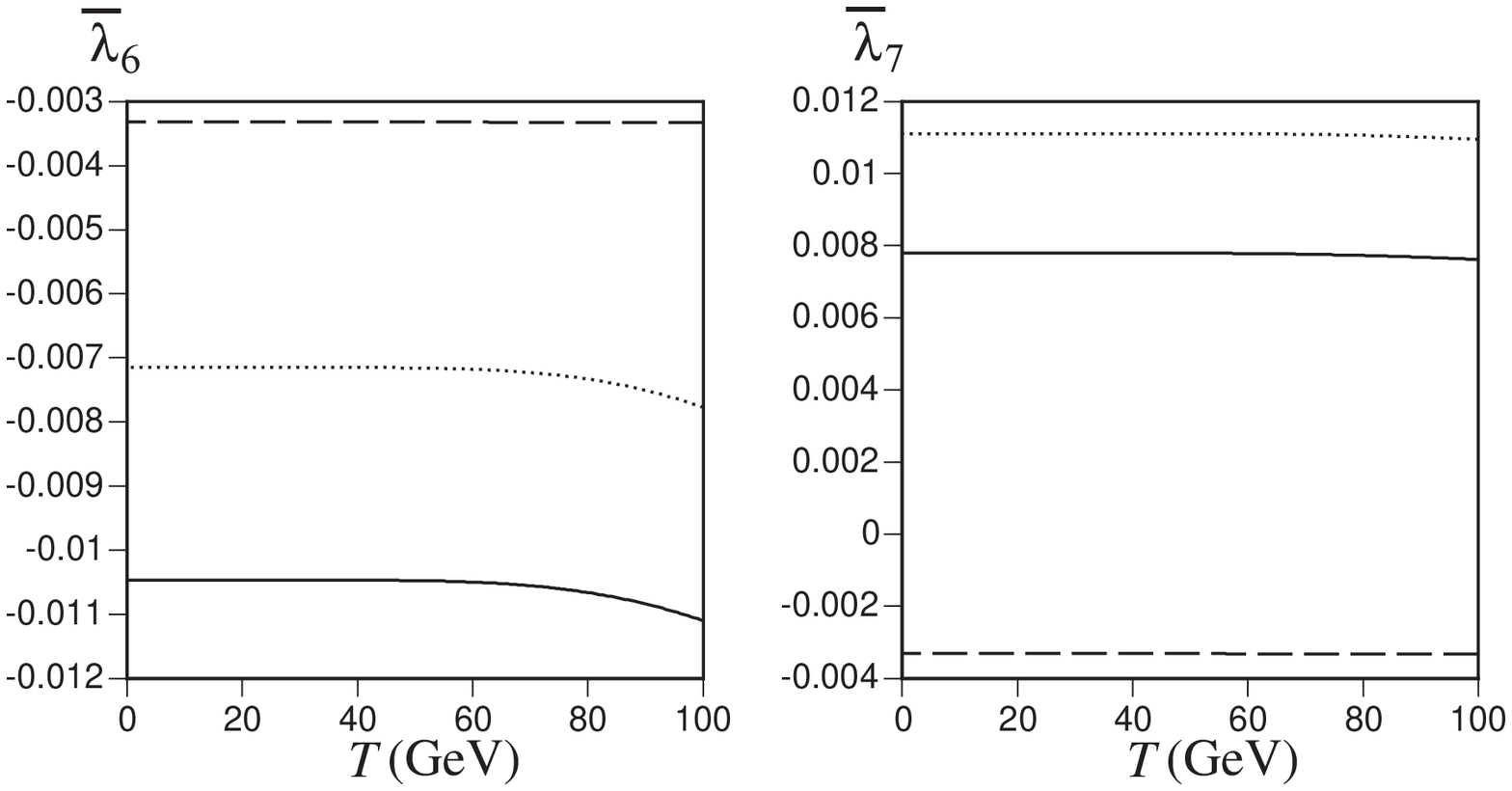}}
 \caption{Behaviors of the effective parameters $\bar\lambda_6$
 and $\bar\lambda_7$. The dashed lines represent the contributions from 
 the charginos and neutralinos, the dotted lines represent those from 
 the stops and the solid lines are the sum of the tree-level and 
 one-loop contributions.}
 \label{fig:eff-para-2}
\end{figure}

\begin{table}
\caption{Parameters adopted for the numerical calculation.}
\label{tab:1}
\begin{center}
\begin{tabular}{cccccccc}
\hline\hline
 $v_0$ & $\tan\beta_0$ & $m_3^2$ &$\mu$  & $A_t$ &$M_2=M_1$&$m_{\st_L}$&$m_{\st_R}$\\
\hline
 $246$ GeV & $6$ &$8110\mbox{ GeV}^2$&$-500$ GeV &$60$ GeV &$500$ GeV &$400$ GeV &$0$\\
\hline
\end{tabular}
\end{center}
\end{table}
The particle spectrum from this set of the parameters are given in
Table~\ref{tab:2}.
\begin{table}
\caption{Masses of the lighter Higgs scalar ($m_h$), the Higgs 
pseudoscalar ($m_A$), the heavier Higgs scalar ($m_H$), the lighter
top squark ($m_{\st_1}$), the lighter chargino ($m_{\chi^\pm_1}$)
and the lightest neutralino ($m_{\chi^0_1}$).}
\label{tab:2}
\begin{center}
\begin{tabular}{cccccc}
\hline\hline
 $m_h$  & $m_A$ &$m_H$ & $m_{\st_1}$ &$m_{\chi^\pm_1}$&$m_{\chi^0_1}$ \\
\hline
 $82.28$ GeV &$117.9$ GeV &$124.0$ GeV &$167.8$ GeV &$457.6$ GeV &
 $449.8$ GeV   \\
\hline
\end{tabular}
\end{center}
\end{table}
The transition temperature $T_{C}$ and the VEVs at $T_{C}$ are given 
by
\begin{equation}
 T_{C}=93.4\mbox{ GeV}, \quad v=129.17\mbox{ GeV},\quad 
 \tan\beta=7.292,
\end{equation}
so that $v/T_{C}\simeq 1.38$, which guarantees that the BAU
created at the EWPT is not washed out after the transition.
\begin{table}
\caption{The effective parameters near the transition temperature 
$T_C=93.4\mbox{ GeV}$.}
\label{tab:3}
\begin{center}
\begin{tabular}{c|cccc}
\hline\hline
 $T$(GeV) &$\left({\bar m}_3^2\right)_{\rm eff}$ ($\mbox{GeV}^2$)&
 $\bar\lambda_5$&$\bar\lambda_6$&$\bar\lambda_7$ \\
\hline
 $93.3$&$-402.375$&$4.98562\times10^{-3}$&$-1.09172\times10^{-2}$&$7.65974\times10^{-3}$\\
 $93.4$&$-401.720$&$4.98684\times10^{-3}$&$-1.09197\times10^{-2}$&$7.65912\times10^{-3}$\\ 
 $93.5$&$-401.065$&$4.98806\times10^{-3}$&$-1.09222\times10^{-2}$&$7.65850\times10^{-3}$\\
 $93.6$&$-400.410$&$4.98928\times10^{-3}$&$-1.09247\times10^{-2}$&$7.65787\times10^{-3}$\\
\hline
\end{tabular}
\end{center}
\end{table}
The values of the effective parameters near $T_{C}$ are presented in 
Table~\ref{tab:3} and 
\begin{equation}
 {B_2\over T_{C}}=1.5741\times10^{-3},\quad {C_1\over T_{C}}=3.2539\times10^{-2}.
\end{equation}
The global structure of $V_{\rm eff}$ at $T_C$ is depicted in 
Fig.~\ref{fig:Veff-contour}.
The $T$ dependence of the effective parameters are plotted 
in Figs.~\ref{fig:eff-para-1} and  \ref{fig:eff-para-2}.
\par
By closely investigating the saddle point between the two degenerate minima,
we find that the height of the saddle point is
$V_{\rm max}=3.6516\times 10^9\mbox{ GeV}^4$,
from which the inverse wall thickness $a$ is given by
\footnote{The solution for $\phi''={{dV_{\rm eff}}\over{d\phi}}$ with 
$V_{\rm eff}={\lambda\over4}\phi^2(\phi-v)^2$ is given by 
$\phi={v\over  2}(1+\tanh az)$ with $a^2={\lambda\over 8}v^2$. 
This leads to 
$V_{\rm max}=V_{\rm eff}(\phi={v\over 2})={\lambda\over{64}}v^4={a^2v^2\over 8}$.}
\begin{equation} 
 a={\sqrt{8V_{\rm max}} \over v}=13.23 \mbox{ GeV}.
\end{equation}
\par
%
%%%%%%%%%%%%%%%%%%%%%%%%%%%%%%%%%%%%%%%%%%%%%%%%%%%%%
\subsection{Transitional $CP$ violation and BAU}\par
\begin{figure}
 \epsfysize=6.0cm
 \centerline{\epsfbox{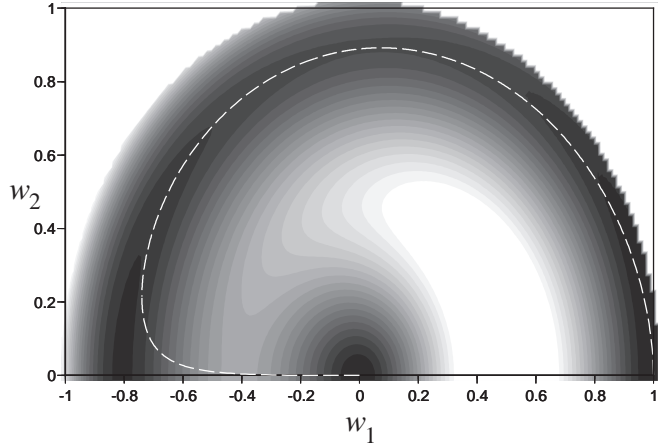}}
 \caption{Contour plot of $W(w_1,w_2)$  together with 
 the positive-$\theta$ solution for $\delta=0.001$ 
 (broken curve).}
 \label{fig:w-cont-1}
\end{figure}

From the numerical results above, we fix the parameters 
at $T_{C}$=93.4 GeV as $v$=129.2 GeV, $a$=13.23 GeV, 
$\tan\beta$=7.29 and $(b=-0.308, c=0.048, d=-0.009, e=-0.3, 
f=-0.0003)$.
For $\delta$=0.001 and 0.002 respectively, a pair of solutions 
$(\rho^{+}(y),\theta^{+}(y))$ and $(\rho^{-}(y),\theta^{-}(y))$ with 
$\theta_{s}=\pi-\delta$ are obtained, which are shown in 
Fig.~\ref{fig:w-cont-1} of 
the $w_{1}$-$w_{2}$ plain on the contour plot of $W(w_1,w_2)$, 
 where 
$w_{1}\equiv\rho\cos(\theta-\theta_{b})$ and 
$w_{2}\equiv\rho\sin(\theta-\theta_{b})$
\footnote{ 
 For $b<0 (>0)$ it is easy to show that the exit gate toward $w_{1}<0$(left)  
 from the basin of the symmetric vacuum at $(w_{1}=0,w_{2}=0)$ is 
 wider (narrower) than that toward $w_{1}>0$ (right) in Fig.~\ref{fig:w-cont-1}. 
 So, $b<0 (>0)$ favors $\theta_{s}=\pi-\delta (=-\delta)$. 
 Usually we have found no convergent solutions with 
 $\theta_{s}=-\delta(\pi-\delta)$
 for  $b<0 (>0)$ in the relaxation method.    
 For $b<0$ with 
 $\tan\beta \leq 3$, no solutions have been obtained. 
 For the smaller values of $\tan\beta$ there is a wide parameter range to admit
 solutions for $b>0$. 
 However, the parameters of them, in particular $(c,d,e)$,  
 are not compatible with the  MSSM calculation. }.
\par\noindent
%%%%%%%%%%%%%%%%%%%%%%%%%%%%%%%%%%%% 
%\vspace{0.5 cm}
{\bf Bubble formation rate}$\qquad$      
 The bubble formation probability of the each solution is proportional to
\begin{equation}
  N^{\pm} =
  \exp\left(-\frac{4\pi R_{crit}^{2}{\Delta{\cal E}^{\pm}}}{T_{C}}\right)
\label{eq:bfp},
\end{equation}
where $R_{crit}$ is the radius of the critical bubble at $T_{C}$ and is given 
by $\sqrt{3F_{crit}/4\pi av^{2}}$ and the free energy of the 
bubble $F_{crit}$ is 
estimated as $145T_{C}$. 
The explicit $CP$ breaking $\delta\neq 0$, even when very small, 
nonperturbatively breaks the degeneracy of the $CP$-conjugate pair of the 
bubbles as mentioned before. 
 The formation rate is
\begin{eqnarray}
    N^{-}/N^{+}&=&0.601\quad     {\rm for}\quad \delta=0.001,\nonumber\\
               &=&0.361\quad     {\rm for}\quad \delta=0.002.
\end{eqnarray}
\par\noindent
{\bf Chiral charge flux }$\qquad$
Once a set of solutions are given, we calculate the reflection 
coefficient $\Delta R(\xi,p_{L}/a)$ of the each solution, where  
$p_{L}$ is the incident momentum of the particle  
perpendicular to the wall and $\xi=m_{C}/a$, $m_C$ being the mass at $T_C$. 
That is, $m_{C}=m_{b(\tau)}v{\cos\beta}/({v_{0}}{\cos\beta_{0}})$ 
for the bottom quark ($\tau$ lepton).
Important here is that the bottom quark and the $\tau$ lepton interact with 
the wall magnitude  
$\rho(y)$  with the coupling $\sin^{2}\beta$ while  the top quark does  
with 
$\cos^{2}\beta$. Therefore the top quark passes the wall almost freely for 
such a large $\tan\beta$ and is completely negligible for the 
BAU.\par\noindent
\par
Then the chiral charge flux through the wall is given by\cite{FKOTTc}
\begin{equation}                        
  F_{Q} =
  \frac{Q_{L}-Q_{R}}{4\pi^{2}\gamma}\int^\infty_{m_C}dp_L
  \int^\infty_0 dp_T\,p_T\Bigr[f^{s}-f^{b}\Bigl]\Delta R(\xi,\frac{p_{L}}{a}),
\end{equation}
where 
\begin{equation}         
   f^{s}=\frac{p_{L}}{E}\frac{1}{{\rm 
   e}^{\gamma(E-up_{L})/T_{C}}+1},\quad             
   f^{b}=\frac{p_{L}}{E}\frac{1}{{\rm e}^{\gamma(E+u\sqrt{p_{L}^{2}-m_{C}^{2}}
)/T_{C}}+1},
\nonumber
\end{equation}      
are the statistical factors in the symmetric and broken
phases respectively, $p_{T}$ is the incident momentum of the particle  
parallel to the wall, $E=\sqrt{p_{L}^{2}+p_{T}^{2}}$, $u$ is the wall velocity
 and 
$\gamma=1/\sqrt{1-u^{2}}$.\par
\begin{figure}
 \epsfxsize=10.0cm
 \centerline{\epsfbox{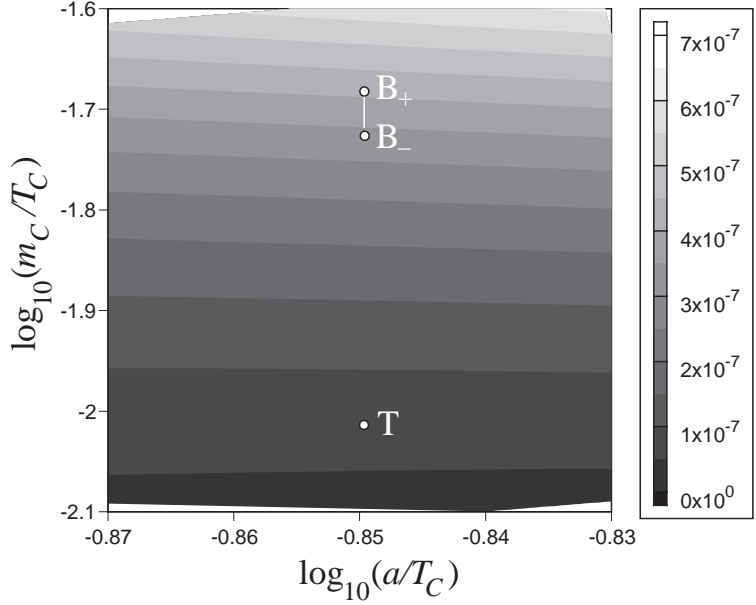}}
 \caption{Contour plot of $F_{Q}^{net}/((Q^{L}-Q^{R})T_{C}^{3}u)$.}
 \label{fig:flux}
\end{figure}

The net contribution to BAU is 
\footnote{
 For the parameter set  with $b<0$, we have found no small-$\theta$
 solutions because the wider exit gate from the 
 symmetric vacuum has the direction just opposite toward the broken 
 one at $(w_{1}=1,w_{2}=0)$. Even in the case when the small solution 
 exists, $\Delta{\cal E}^{s}$ is negligibly small.}
\begin{equation} 
   F_{Q}^{net}=
   {{N^{+}F_{Q}^{+}-N^{-}F_{Q}^{-}}\over{N^{+}+N^{-}}}.
\end{equation}
Fig.~\ref{fig:flux} shows $F_{Q}^{net}/((Q^{L}-Q^{R})T_{C}^{3}u)$ 
for $\delta=0.001$ and $u=0.1$ as a contour plot on 
 the $a/T_{C}-m_{C}/T_{C}$ plain. The point ${\rm B_{+}(B_{-})}$ is the 
$b$ quark  
contribution if $m$=4.4(4.1) GeV, and the point T is the 
$\tau$ lepton contribution. 
 $F_{Q}^{net}/((Q^{L}-Q^{R})T_{C}^{3}u)$ amounts to $4\times 10^{-7}, 
3.4\times 10^{-7},4\times 10^{-8}$ at ${\rm B_{+}, B_{-}}$, T 
respectively. 
 The top of the contour hill is $6.3\times 
10^{-3}$ at $\log_{10}(a/T_{C})=0.88$ and 
$\log_{10}(m_{C}/T_{C})=0.29$. As is well known, a thin wall 
 produces a large chiral charge flux.   For $\delta=0.002$, $F_{Q}^{net}$
increases as a whole by a factor of about $1.7$. As $u$ increases to 
0.9, the contour at the point $B_{-}$, say,  decreases by a factor of 
about  0.4.\par
% {\bf Fig.7}. Contour plot of 
% $F_{Q}^{net}/((Q^{L}-Q^{R})T_{C}^{3}u)$.\par
%
From $F_Q^{net}$ we can estimate the BAU  in the transport scenario as
\begin{equation}
 {\rho_B\over s}\sim 10^{-7}\times{F_Q^{net}\over u}\times{\tau\over T_{C}^2},
\end{equation}
where the entropy density is given by $s=2\pi^2g_*T_{C}^3/45$ with $g_*\simeq1
00$ 
and $\tau$ is the transverse time within which the scattered particles are 
captured by the wall.
If $\tau$ is given by $D/u$ with the diffusion constant $D$, it is 
estimated from $D(b\quad {\rm quark})\sim1/T_{C}$ and $D(\tau\quad {\rm lepton
})\sim
(10^{2-3})/T_{C}$. Then
\begin{eqnarray}
\rho_B/s &<& 10^{-12} \quad
({\rm for}\quad b\quad {\rm quark}),  \nonumber\\
\rho_B/s &\sim& 10^{-(10-12)} \quad
({\rm for}\quad \tau\quad {\rm lepton})
\end{eqnarray}
for $\delta=0.001$ and $u=0.1$. Thus our example may give a possibility to
produce the BAU by the transitional $CP$ violation of the $\tau$ lepton, if 
marginally, in the MSSM, provided that $\delta\sim O(10^{-3})$ is induced.
%
%
%%%%%%%%%%%%%%%%%%%
\section{Summary and Discussions}
We have shown the possibility of the transitional $CP$ violation 
in the MSSM. 
For this scenario to be realized, some of the parameters in the theory
are constrained: $\mu M_2<0$ and/or $\mu A_t<0$ are required, 
$m_3^2$ should be $O(10^4)\mbox{ GeV}^4$, and $\tan\beta_0\ge 5$
would be necessary for our choice of the mass parameters to
have the lightest Higgs scalar with an acceptable mass and to
have nontrivially $CP$-violating wall profile.
These requirements inevitably relate the mass of the Higgs scalar 
to that of the pseudoscalar.\par
Although the explicit $CP$ breaking in the Higgs sector is severely 
limited from the neutron EDM  at $T=0$, that induced at $T_C$ can be 
somewhat large. If the latter is $\delta\sim O(10^{-3})$ and breaks the 
degeneracy between the pair of $CP$-conjugate bubbles, our numerical example 
could be able to produce the BAU$\sim O(10^{-10})$ by the $\tau$ lepton 
transport.
The $\theta$-dependent $\rho^3$-terms essential for the transitional $CP$ 
violation is induced only if one of the soft SUSY-breaking masses, 
$m_{\tilde t_R}$ of the stop, almost vanishes.
This implies that the mass of the lighter stop $m_{\st_1}$ must 
satisfy $m_{\st_1}\ltsim m_t$.\par
Our result of a rather thick walls, $1/a\sim 10/T_{C}$ together 
with a large $\Delta \theta=|\theta_{b}-\theta_{s}| \sim \pi$ may favor 
the diffusion scenario\cite{diffusion}, 
although estimating various uncertain factors of  this 
scenario is outside the scope of this article.\par
\begin{figure}[h]
 \epsfysize=6.0cm
 \centerline{\epsfbox{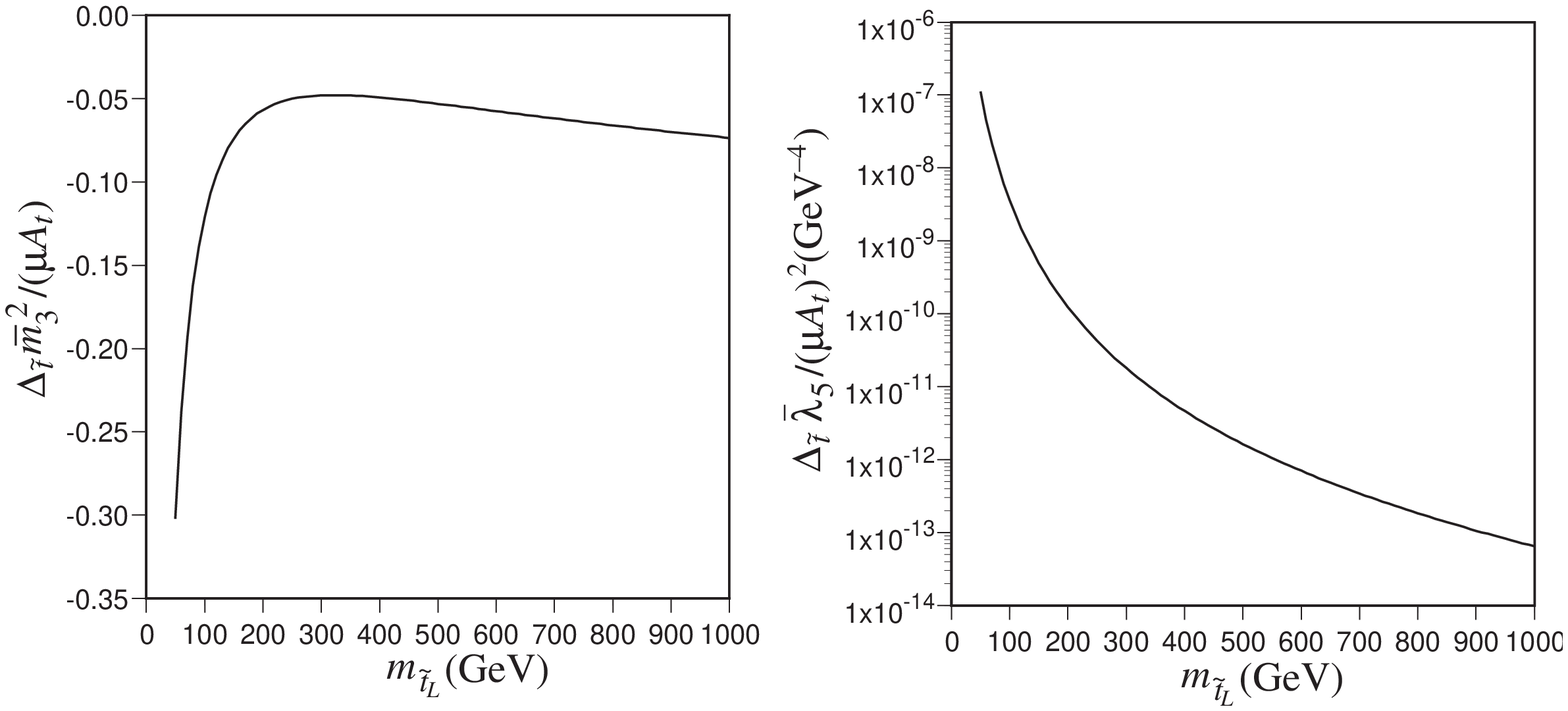}}
 \caption{$\Delta_{\st}m_3^2/(\mu A_t)$ and 
 $\Delta_{\st}\lambda_5/(\mu A_t)^2$ for $\tan\beta_0=6$ 
 at $T=95\mbox{ GeV}$.}
 \label{fig:stop-contribution}
\end{figure}
\begin{figure}[h]
 \epsfysize=6.5cm
 \centerline{\epsfbox{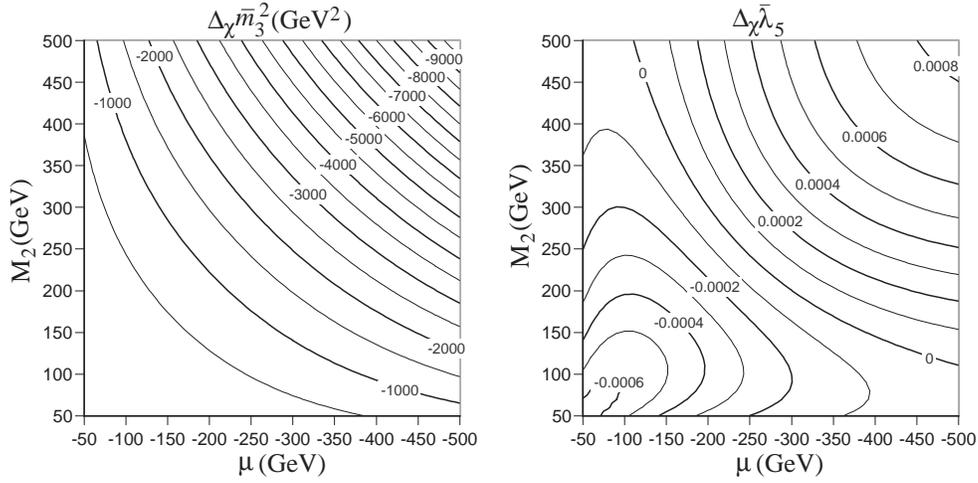}}
 \caption{Contour plots of $\Delta_{\chi}m_3^2$ and
 $\Delta_\chi\lambda_5$ at $T=95\mbox{ GeV}$.}
 \label{fig:chargino-contribution}
\end{figure}

In \S 2.2, we have pointed out another possibility of transitional
$CP$ violation associated with $F(\rho_1,\rho_2)<0$.
A wall profile of this type, if exists, will generate sizable
BAU just as the example presented here, since $\theta$ varies
from $0$ to $\pi$.
As long as the high-temperature expansion of the light stop 
contributions is valid, the $B_2$-term in (\ref{eq:def-F}) is negligible 
compared to the $\bar\lambda_5$-term.
Indeed we found a well qualified solutions for $(d=-0.009,f=-0.0003)$,
which is obtained from the parameter set studied here only by
changing the sign of $d$.
It is, however, difficult to have negative $\bar\lambda_5$ and
$\abs{G(\rho_1,\rho_2)}<1$ for an intermediate $(\rho_1,\rho_2)$
with an acceptable set of parameters when $m_{\st_R}=0$.
This can be seen as follows. As shown in \cite{FKOTd}, 
$\Delta_\st m_3^2/(\mu A_t)$ and $\Delta_\st\lambda_5/(\mu A_t)^4$
are functions of $T$ and $m_{\st_L}$ when $m_{\st_R}=0$.
These are plotted for $\tan\beta_0=6$ at $T=95\mbox{ GeV}$ 
in Fig.~\ref{fig:stop-contribution}.
Contour plots of the chargino-neutralino contributions 
$\Delta_\chi m_3^2$ and $\Delta_\chi\lambda_5$ for
$\tan\beta_0=6$ and $T=95\mbox{ GeV}$ are shown in
Fig.~\ref{fig:chargino-contribution}.
In order for $\bar\lambda_5$ to be negative, rather small
$\abs{\mu M_2}$ is required. Since $\abs{\Delta_\chi\lambda_5}$
is at most $O(10^{-4})$ in the region where $\Delta_\chi\lambda_5$
is negative, $\abs{\mu A_t}$ must be smaller than $10^4\mbox{ GeV}^2$ 
to have negative $\bar\lambda_5$. As seen from 
Fig.~\ref{fig:chargino-contribution}, 
$\Delta_\chi m_3^2>1500\mbox{ GeV}^2$ in the region where 
$\Delta_\chi\lambda_5<0$. As long as we take $m_{\st_L}$ larger
than the weak scale, $\Delta_\st m_3^2\sim -0.1\cdot \mu A_t$,
so that $\Delta_\st m_3^2$ should be larger than $-1000\mbox{ GeV}^2$.
Hence to satisfy $\abs{G(\rho_1,\rho_2)}<1$, the tree-level
$m_3^2$ must be taken to be at most $2500\mbox{ GeV}^2$.
This small $m_3^2$ inevitably produces too light Higgs bosons,
which are inconsistent with the present bound $m_h\ge 67.5\mbox{ GeV}$
and $m_A\ge 67.5\mbox{ GeV}$\cite{OPAL}.
We have examined cases with larger $\tan\beta_0$, which yield
heavier Higgs bosons, but find no region in the parameter space
that both $\bar\lambda_5<0$ and $\abs{G(\rho_1,\rho_2)}<1$
are satisfied with acceptable Higgs masses.
Although this interesting scenario of $CP$ violation is far from
realizable in the MSSM, 
a general two-Higgs-doublet model will allow a wall profile with
$\bar\lambda_5<0$, since it has nonzero tree-level $\lambda_5$
as a free parameter and we confirmed the existence of such a
solution.
\begin{center}
{\bf Acknowledgment}
\end{center}
The authors cordially express their gratitude to A.~Kakuto for his
valuable discussions and enlightening comments.
This work is supported in part by a Grand-in-Aid for Scientific Research on
 Priority Areas(Physics of $CP$ violation, No.~10140220), No.~09740207 (K.F.) 
and No.~09640378 (F.T.) from the Ministry of Education,Science and Culture 
of Japan.
%%%%%%%%%%%%%%%%%%%%%%%%%%%%%
%
%
\appendix  
\section{Formulas for the Effective Parameters}
In this appendix, we summarize the contributions from the charginos,
neutralinos and stops to the effective parameters.
As for the formulas to evaluate the finite-temperature Feynman 
integrals, refer to Appendix of \cite{FKOTd}.\par
The mass matrices of the charginos and neutralinos are given by
\begin{eqnarray}
 M_{\chi^{\pm}} &=& \left(
            \begin{array}{cc}
            M_2 & -{ig_2\over\sqrt2}\rho_2 e^{-i\theta} \\
            -{ig_2\over\sqrt2}\rho_2  & -\mu 
           \end{array}
           \right),   \\
 M_{\chi^0} &=& \left(\begin{array}{cccc}
 M_2 & 0 & -{i\over 2}g_2\rho_1 & {i\over 2}g_2\rho_2 e^{-i\theta} \\
 0 & M_1 & {i\over 2}g_1\rho_1 & -{i\over 2}g_1\rho_2 e^{-i\theta} \\
 -{i\over 2}g_2\rho_1 & {i\over 2}g_1\rho_1 & 0 & \mu \\
{i\over 2}g_2\rho_2e^{-i\theta} & -{i\over 2}g_1\rho_2e^{-i\theta}& \mu & 0 \\
           \end{array}\right), 
\end{eqnarray}
respectively. Here $M_1$ and $M_2$ are the gaugino mass parameters.
The contributions from the charginos are given by
\begin{eqnarray}
 \Delta_{\chi^\pm} m_3^2 &=&
 2 g_2^2 \mu M_2\cdot i\int_k\Delta_1(k)\Delta_\mu(k),
        \label{eq:m32-chargino}\\
 \Delta_{\chi^\pm}\lambda_5 &=&
 -2g_2^4(\mu M_2)^2\cdot i\int_k\Delta_1^2(k)\Delta_\mu^2(k),
        \label{eq:l5-chargino}\\
 \Delta_{\chi^\pm}\lambda_6 &=& \Delta_{\chi^\pm}\lambda_7=
 2g_2^4 \mu M_2\cdot i\int_k k^2\Delta_1^2(k)\Delta_\mu^2(k),
        \label{eq:l67-chargino}
\end{eqnarray}
and those from the neutralinos are
\begin{eqnarray}
 \Delta_{\chi^0} m_3^2 &=&
 2 i\int_k\left[g_2^2\mu M_2\Delta_1(k)\Delta_\mu(k) +
                g_1^2\mu M_1\Delta_2(k)\Delta_\mu(k) \right],
        \label{eq:m32-neutralino-0}\\
 \Delta_{\chi^0}\lambda_5 &=&
 -2 i\int_k\left[ g_2^2\mu M_2\Delta_1(k) + g_1^2\mu M_1\Delta_2(k)
           \right]^2 \Delta_\mu^2(k),
        \label{eq:l5-neutralino-0}\\
 \Delta_{\chi^0}\lambda_6 &=&
 2 i\int_k k^2\left[ g_2^4 M_2\Delta_1^2(k) +
   g_2^2g_1^2(M_2+M_1)\Delta_1(k)\Delta_2(k) + g_1^4 M_1\Delta_2^2(k)
              \right] \mu\Delta_\mu^2(k) \nonumber\\
 &=& \Delta_{\chi^0}\lambda_7,
        \label{eq:l67-neutralino-0}
\end{eqnarray}
where
\begin{equation}
 \Delta_1(k) = {1\over{k^2-|M_2|^2}},\qquad
 \Delta_2(k) = {1\over{k^2-|M_1|^2}},\qquad
 \Delta_\mu(k) = {1\over{k^2-|\mu|^2}}.
\end{equation}
Note that $M_2$, $M_1$ and $\mu$ are complex.\par
In the special case of $M_2=M_1$, the corrections to the effective parameters
are reduced to
\begin{eqnarray}
 \Delta_{\chi^0}m_3^2 &=&
 2{{g_2^2}\over{\cos^2\theta_W}}\mu M_2\cdot i\int_k
           \Delta_1(k)\Delta_\mu(k),    \label{eq:m32-neutralino}\\
 \Delta_{\chi^0}\lambda_5 &=&
 -2{{g_2^4}\over{\cos^4\theta_W}}\left(\mu M_2\right)^2\cdot
    i\int_k \Delta_1^2(k)\Delta_\mu^2(k),  \label{eq:l5-neutralino}\\
 \Delta_{\chi^0}\lambda_6 &=& \Delta_{\chi^0}\lambda_7 =
 2{{g_2^4}\over{\cos^4\theta_W}}\mu M_2\cdot
    i\int_k k^2\Delta_1^2(k)\Delta_\mu^2(k).\label{eq:l67-neutralino}
\end{eqnarray}
In this case, the contributions of charginos relate to those of neutralinos as
\begin{eqnarray}
 \Delta_{\chi^0}m_3^2 &=&
 {1\over{\cos^2\theta_W}} \Delta_{\chi^\pm}m_3^2,         \\
 \Delta_{\chi^0}\lambda_5 &=&
 {1\over{\cos^4\theta_W}} \Delta_{\chi^\pm}\lambda_5,     \\
 \Delta_{\chi^0}\lambda_6 &=& \Delta_{\chi^0}\lambda_7 =
 {1\over{\cos^4\theta_W}} \Delta_{\chi^\pm}\lambda_6 =
 {1\over{\cos^4\theta_W}} \Delta_{\chi^\pm}\lambda_7.
\end{eqnarray}
\par\noindent
The mass-squared matrix of stops is given by
\begin{equation}
 M_{\tilde t}^2 = \pmatrix{ m_{11}^2 & m_{12}^2 \cr
                     m_{12}^{2*} & m_{22}^2 \cr}, \label{eq:def-Mt}
\end{equation}
where
\begin{eqnarray}
 m_{11}^2&=&
  m_{\tilde t_L}^2-{1\over8}\left({{g_1^2}\over3}-g_2^2\right)
  (\rho_1^2-\rho_2^2)+\half y_t^2\rho_2^2,  \label{eq:def-m11}\\
 m_{22}^2&=&
  m_{\tilde t_R}^2+{1\over6}g_1^2(\rho_1^2-\rho_2^2)+\half y_t^2\rho_2^2,
                                                  \label{eq:def-m22}\\
 m_{12}^2&=&
  {{y_t}\over{\sqrt{2}}}\left[(\mu \rho_1+A_t\rho_2\cos\theta)-
 iA_t\rho_2\sin\theta\right].
                                                  \label{eq:def-m12}
\end{eqnarray}
Here $y_t$ is the top Yukawa coupling. $m_{\tilde t_L}^2$, 
$m_{\tilde t_R}^2$ and $A_t$ are the soft-SUSY-breaking mass-squared parameters.\par 
The formulas for the corrections to the effective parameters are
\begin{eqnarray}
 \Delta_\st m_3^2 &=&
 -N_c y_t^2\mu A_t^*\cdot i\int_k \Delta_1(k)\Delta_2(k),
                     \label{eq:m32-stop}\\
 \Delta_\st\lambda_5 &=&
 N_c y_t^4(\mu A_t^*)^2\cdot i\int_k \Delta_1^2(k)\Delta_2^2(k),
                     \label{eq:l5-stop}\\
 \Delta_\st\lambda_6 &=&
 -N_c y_t^2\mu A_t^*\cdot i\int_k\Biggl[
   \left({{g_2^2}\over4}-{{g_1^2}\over{12}}\right)\Delta_1^2(k)\Delta_2(k)
  +{{g_1^2}\over3}\Delta_1(k)\Delta_2^2(k)  \nonumber\\
 & &\qquad\qquad\qquad\qquad\qquad
 +y_t^2|\mu|^2\Delta_1^2(k)\Delta_2^2(k) \Biggr],\label{eq:l6-stop}\\
 \Delta_\st\lambda_7 &=&
 -N_c y_t^2\mu A_t^*\cdot i\int_k\Biggl\{
   \left[y_t^2-\left({{g_2^2}\over4}-{{g_1^2}\over{12}}\right)\right]
      \Delta_1^2(k)\Delta_2(k)
  +\left(y_t^2-{{g_1^2}\over3}\right)\Delta_1(k)\Delta_2^2(k)  \nonumber\\
 & &\qquad\qquad\qquad\qquad\qquad
 +y_t^2|A_t|^2\Delta_1^2(k)\Delta_2^2(k) \Biggr\},\label{eq:l7-stop}
\end{eqnarray}
where
\begin{equation}
 \Delta_1(k) = {1\over{k^2-m_{\tilde t_L}^2}},\qquad
 \Delta_2(k) = {1\over{k^2-m_{\tilde t_R}^2}}.
\end{equation}
In the case of $m_{\st_R}\simeq0$, the finite-temperature Feynman 
integrals become ill-defined because of infrared divergence. Then
one should employ another method to evaluated the effective potential
as done in \cite{FKOTd}.
$B_2$ and $C_1$ are extracted from the light stop contribution to the
effective potential, which is treated by the high-temperature expansion
(\ref{eq:high-T-exp-J}).
%
%
%%%%%%%%%%%%%%%%%%%%%%%%%%%%%%%%%%%%%%

%
%
%
%
\end{document}